\documentclass[12pt,a4paper,aps,prd,preprint,superscriptaddress,nofootinbib]{revtex4-1}
\usepackage[utf8]{inputenc}
\usepackage{graphicx}
\usepackage{amssymb}
\usepackage{textcomp}
\usepackage{amsmath}
\usepackage{tabularx}
\usepackage{bm}
\usepackage{times}
\usepackage{color}
\usepackage{slashed}
\usepackage{multirow}
\usepackage{verbatim}
\usepackage{cancel}
\usepackage{subfigure}
\usepackage[normalem]{ulem}
\usepackage{float}

\usepackage[colorlinks=true, pdfstartview=FitV, linkcolor=blue, citecolor=blue, urlcolor=blue]{hyperref}
\allowdisplaybreaks[4]

\def\lsim{\mathrel{\raise.3ex\hbox{$<$\kern-.75em\lower1ex\hbox{$\sim$}}}}
\def\gsim{\mathrel{\raise.3ex\hbox{$>$\kern-.75em\lower1ex\hbox{$\sim$}}}}

\definecolor{orange}{rgb}{1,0.5,0}

\begin{document}

\title{Constraint on ultralight Nelson-Barr dark matter from time-dependent nuclear decay}

\author{Chang-Jie Dai}
\email{daichangjie@mail.nankai.edu.cn}
\affiliation{
School of Physics, Nankai University, Tianjin 300071, China
}

\author{Tong Li}
\email{litong@nankai.edu.cn}
\affiliation{
School of Physics, Nankai University, Tianjin 300071, China
}

\begin{abstract}
Many experiments have notably reported anomalies in radioactive decay rates with periodic variations and challenged the traditional belief of time-independent decays. This periodicity could potentially be explained by the presence of a periodic dark matter (DM) candidate. In this work, we investigate the impact of time-dependent nuclear decay rates on the ultralight scalar DM in Nelson-Barr solution to the strong CP problem. The light scalar DM field in this framework induces periodic modulations of both CKM matrix elements and quark mass parameters. These modulations generate corresponding periodic perturbations in the neutron-proton mass difference and nuclear binding energies. Consequently, nuclear decay rates receive oscillating residuals. By analyzing the tritium decay rate, we derive new exclusion bounds on the Nelson-Barr DM parameters.
\end{abstract}

\maketitle

\tableofcontents

\newpage

\section{Introduction}

Early observations of radioactive decay in the era of M.~Curie support a long-standing perspective that nuclear radioactivity is independent of time. The number of remaining nuclei follows $N(t)=N_0 e^{-\lambda t}$ where the probability of decay per unit time $\lambda$ was traditionally considered constant over time. Indeed, under normal conditions (room temperature, standard pressure, no extreme environments), radioactive isotopes (e.g., uranium-238, carbon-14) exhibit highly consistent decay rates over human timescales.
It attributes to a statistical nature that individual nuclei decay randomly, but the ensemble behavior appears time-independent.

Numerous subsequent experiments have notably reported anomalies in radioactive decay rates with periodic variations and challenged the traditional belief of time-independent decay constants (see bibliography~\cite{McDuffie:2020uuv} and references therein). This periodicity could potentially be explained by the annual variation in solar neutrino flux~\cite{Davis:1995wj,Jenkins:2008tt,Mohsinally:2016juj} or the presence of a periodic dark matter (DM) candidate~\cite{Agafonova:2021tgr}.
However, the evidence of this annual periodicity remains under debate due to experimental influences and uncertainty analysis~\cite{JRC90676,Pomme:2016yoh,Pomme_2017I,Pomme_2017II,Pomme_2017III,Pomme:2018vgo,Pomme:2019mhq,Pomme:2020veo}.
Indeed, some more recent measurements and statistical analysis excluded modulations of the decay constants in tritium~\cite{Zhang:2023lem} and several heavier radioisotopes~\cite{Bellotti:2012if,Bellotti:2013bka,Bellotti:2015toa,Bellotti:2018jzd}.

The null significant observation of periodic variations in nuclear decay rates allows us to place constraints on potential time-varying nature of physics beyond the Standard Model (SM). A well-motivated new physics with time-dependence is the QCD axion field as the consequence of strong CP problem solution~\cite{Peccei:1977hh,Peccei:1977ur,Weinberg:1977ma,Wilczek:1977pj} and nature DM candidate (see a recent review Ref.~\cite{DiLuzio:2020wdo}). Several recent studies have investigated nuclear decays as a probe of axion DM and established new constraints on the axion decay constant in the ultralight mass range~\cite{Zhang:2023lem,Broggini:2024udi,Alda:2024xxa}.

The Nelson-Barr model raises the possibility of spontaneous breaking of CP symmetry and provides another solution to the strong CP problem~\cite{Nelson:1983zb,Nelson:1984hg,Barr:1984qx}. This model postulates exact CP symmetry at high-energy scale, and introduces additional vector-like fermions coupled to SM quarks and a complex scalar field $\Phi$~\cite{Bento:1991ez}. The spontaneous symmetry breaking of $\Phi$ generates an irreducible CP-violating phase within the Cabibbo-Kobayashi-Maskawa (CKM) matrix, thereby resolving the strong CP problem without introducing nonzero $\bar{\theta}_{\rm QCD}$. This mechanism confines CP violation exclusively to observable weak interactions while preserving the natural vanishing of $\bar{\theta}_{\rm QCD}$
at the fundamental level. The pseudo-Nambu Goldstone boson associated with spontaneously broken CP symmetry emerges as a viable candidate for ultralight scalar DM~\cite{Dine:2024bxv} (see e.g. Ref.~\cite{Khlopov:1985fch} for other ultralight scalar DM). Notably, such a scalar field $\phi$, when constituting the DM background, induces temporal oscillations in the CKM matrix elements through its coherent cosmic oscillations. These field-dependent modulations of weak interaction parameters consequently result in periodic modifications to effective quark masses via Yukawa coupling renormalization and establish a characteristic oscillating signature within the SM fermion sector.

The periodic oscillations of quark masses propagate into modulations of both the neutron-proton mass difference and nuclear binding energies through isospin-dependent nucleon interactions. Previous studies have established that axion DM could induce periodic variations in nuclear decay rates via analogous mechanisms affecting these nuclear parameters. In this work, we extend this paradigm to $\phi$-dependent modulations within the Nelson-Barr framework. By implementing a frequentist hypothesis-testing protocol on precision measurements of tritium $\beta$-decay rate, we derive novel constraints on the DM particle mass $m_{\phi}$
and decay constant $f$.

This paper is organized as follows.
In Sec.~\ref{sec:NB}, we review the Nelson-Barr model and the ultralight scalar DM therein.
We present the detailed calculations of nuclear decay rate in this framework in Sec.~\ref{sec:Calc}. The constraint on Nelson-Barr DM is then shown in Sec.~\ref{sec:Cons}. Our conclusions are drawn in Sec.~\ref{sec:Con}.

\section{Nelson-Barr model and ultralight scalar dark matter}
\label{sec:NB}

In Nelson-Barr model, one introduces a vector-like up-type~\footnote{Here we adopt the framework in Ref.~\cite{Dine:2024bxv}. The conventional Nelson-Barr studies instead introduced a vector-like down-type quark~\cite{Bento:1991ez,Dine:2015jga,Valenti:2021rdu}.} quark $\mathcal{U}\sim (3,1,2/3)$ and a complex scalar singlet $\Phi$~\cite{Dine:2024bxv}. The renormalizable Lagrangian for the quark sector reads
\begin{eqnarray}
-\mathcal{L}\supset \mu~\overline{\mathcal{U}_L}\mathcal{U}_R+(g_i\Phi+\tilde{g}_i \Phi^\ast)\overline{\mathcal{U}_L} u_{Ri}+y^u_{ij}\tilde{H}\overline{Q_{Li}}u_{Rj}+y^d_{ij}H\overline{Q_{Li}}d_{Rj}+h.c.\;,
\end{eqnarray}
where $H$ denotes the SM Higgs doublet, $\tilde{H}=i\sigma_2 H^\ast$ and $\Phi=(\rho+f)e^{i\theta}/\sqrt{2}$. The interactions $\tilde{H}\overline{Q_{L}}\mathcal{U}_{R}$ and $\Phi \overline{\mathcal{U}_L}\mathcal{U}_R$ are forbidden by introducing a $Z_2$ symmetry under which the new fields are odd and the SM fields are even.
After electroweak symmetry breaking, the Higgs field $H$ gains vacuum expectation value $v$ and the mass matrix of up-type quarks becomes
\begin{align}
\mathcal{M}_u=\left(
\begin{array}{cc}
	\mu & B \\
	0 & m_u \\
\end{array}
\right)\;,
\end{align}
where $m_u=y^uv/\sqrt{2}$ and $B_i=g_i\langle \Phi\rangle + \tilde{g}_i \langle \Phi^\ast \rangle$.
Assuming $\mu, |B|\gg m_u$, we can diagonalize the block matrix $\mathcal{M}_u$ and get the new mass matrix for the light up-type quarks satisfying
\begin{eqnarray}
\tilde{m}_u \tilde{m}_u^\dagger = m_u m_u^T - {(m_u B^\dagger) (B m_u^T)\over \mu^2+BB^\dagger}\;.
\label{effective mass matrix}	
\end{eqnarray}
We take the basis in which the original up-type quark mass matrix $m_u$ is diagonal and the down-type quark mass matrix is
\begin{eqnarray}
m_d=y^d v/\sqrt{2}\;,~~~V^{d\dagger}m_d m_d^\dagger V^d = m_d^{\rm diag} m_d^{{\rm diag}T}\;,
\end{eqnarray}
where the unitary transformation matrix $V^d$ becomes the original CKM matrix.

Following Ref.~\cite{Dine:2024bxv}, we choose non-zero components of $g_i$ and $\tilde{g}_i$ as
\begin{eqnarray}
g_1\neq 0\;,~~\tilde{g}_2\neq 0\;.
\end{eqnarray}
A rotation $O_{12}$ is then introduced to diagonalize the first two up-type quarks in $\tilde{m}_u$ and the phase in $\Phi$ can be removed by a phase transformation $P$ as follows
\begin{align}
O_{12}=	
\begin{pmatrix}
\cos \theta_{12}&\sin \theta_{12}&0\\
-\sin \theta_{12}&\cos \theta_{12}&0\\
0&0&1
\end{pmatrix}\;,~~~
P=
\begin{pmatrix}
1&0&0\\
0&e^{-2i\theta}&0\\
0&0&1
\end{pmatrix}\;,
\end{align}
where
\begin{eqnarray}
\tan2\theta_{12}={-2a^2 m_u m_c |g| |\tilde{g}|\over m_u^2(1-a^2 |g|^2)-m_c^2(1-a^2 |\tilde{g}|^2)}\;,~~~a^2={f^2\over 2\mu^2+f^2(|g|^2+|\tilde{g}|^2)}\;.
\end{eqnarray}
We then have the transformation of up-type quark mass matrix as
\begin{eqnarray}
V^{u\dagger}\tilde{m}_u\tilde{m}_u^\dagger V^u = \tilde{m}_u^{\rm diag} \tilde{m}_u^{{\rm diag}T}\;,~~V^u=(O_{12}P)^\dagger\;.
\end{eqnarray}
The physical up and charm quark masses are
\begin{eqnarray}
\tilde{m}_{c,u}^2={1\over 2}\Big[M^2\pm \sqrt{M^{\prime 4}-4m_u^2 m_c^2 (1-a^2 |g|^2 - a^2 |\tilde{g}|^2)} \Big]\;,
\end{eqnarray}
where $M^2=m_u^2(1-a^2 |g|^2)+m_c^2(1-a^2 |\tilde{g}|^2)$ and $M^{\prime 4}=m_u^4(1-a^2 |g|^2)^2+m_c^4(1-a^2 |\tilde{g}|^2)^2$.
The new CKM matrix becomes
\begin{eqnarray}
V=V^{u\dagger}V^d=O_{12}P V^d\;.
\end{eqnarray}
It turns out that the complex scalar phase $\theta$ becomes the origin of CKM phase. Similar to the relaxion framework~\cite{Davidi:2017gir}, a light scalar field can be introduced from the replacement
\begin{eqnarray}
\theta\to \theta_0 + \phi/f\;,
\end{eqnarray}
where $\theta_0$ denotes a background value and the scalar field $\phi$ may constitute DM if it is light enough~\cite{Dine:2024bxv}.
The CKM matrix elements read
\begin{align}
V_{uj}&=V^{d}_{uj}\left(\cos{{\theta}_{12}}+\frac{V^{d}_{cj}}{V^{d}_{uj}}\sin{{\theta}_{12}}\exp\left[-2i\left(\theta_0+\frac{\phi}{f}\right)\right]\right)\,,\\
V_{cj}&=V^{d}_{cj}\left(\cos{{\theta}_{12}}\exp\left[-2i\left(\theta_0+\frac{\phi}{f}\right)\right]-\frac{V^{d}_{uj}}{V^{d}_{cj}}\sin{{\theta}_{12}}\right)\,,
\end{align}
where $j=d,s,b$.
Their absolute values are~\cite{Dine:2024bxv}
\begin{eqnarray}
|V_{u/cj}|^2&=&|V^d_{u/cj}\cos\theta_{12}|^2  \nonumber\\
&\times& \Big[1+\Big({V^d_{c/uj}\over V^d_{u/cj}}\Big)^2\tan^2\theta_{12}\pm 2{V^d_{c/uj}\over V^d_{u/cj}}\tan\theta_{12} \cos\Big(2\theta_0+2{\phi\over f}\Big) \Big]\;.
\label{eq:V2}
\end{eqnarray}
In the limit of $\phi/f\ll 1$, they have the following approximate form
\begin{align}
\label{CKM phi}
|V_{u/cj}|^2\approx&\mp2V^{d}_{u/cj}V^{d}_{c/uj}\kappa\phi/f \notag \\
&+|V^{d}_{u/cj}|^2\times
\left(\cos^2{{\theta}_{12}}+\left(\frac{V^{d}_{c/uj}}{V^{d}_{u/cj}}\right)^2\sin^2{{\theta}_{12}}{\pm}\frac{V^{d}_{c/uj}}{V^{d}_{u/cj}}\kappa  \right)\,,
\end{align}
where $\kappa=\sin 2\theta_{12}\sin2\theta_0$. In the following analysis, we follow Ref.~\cite{Dine:2024bxv} to take $\sin 2\theta_0\sim\cos2\theta_0 \sim 1$, $\theta_{12}\sim \lambda_{\rm Cabibbo}\sim 0.1$ and $\kappa\sim 0.2$.

\section{Calculation of oscillating nuclear decay rate}
\label{sec:Calc}

As established in the previous section, the Nelson-Barr model modifies the CKM matrix by introducing a scalar field $\phi$ from spontaneous CP symmetry breaking. The scalar field $\phi$ is postulated in Ref.~\cite{Dine:2024bxv} as a viable DM candidate and exhibits temporal oscillation
\begin{align}
\frac{\phi(t)}{f}=\frac{\sqrt{2\rho_{\rm DM}}}{f m_{\phi}}\cos (m_{\phi}t)\;,
\end{align}
where $\rho_{\rm DM}=0.45~{\rm GeV}/{\rm cm}^{3}$~\cite{Turner:1990qx,ADMX:2021nhd} denotes the local DM density.
As a result of Eq.~(\ref{CKM phi}), the CKM matrix elements exhibit time-dependent oscillations due to the oscillating field $\phi$
\begin{align}
\label{CKM dependence}
|V_{u/cj}|^2\approx |V_{u/cj}^0|^2 \big( 1\mp 2  T_{u/cj}\kappa\frac{\phi}{f}   \big)~,
\end{align}
where
\begin{align}
\label{CKM}
|V_{u/cj}^0|^2&=|V^{d}_{u/cj}|^2\times
\left(\cos^2{{\theta}_{12}}+\left(\frac{V^{d}_{c/uj}}{V^{d}_{u/cj}}\right)^2\sin^2{{\theta}_{12}}{\pm}\frac{V^{d}_{c/uj}}{V^{d}_{u/cj}}\kappa  \right)\;,~~
T_{u/cj}=\frac{V^{d}_{u/cj}V^{d}_{c/uj}}{|V_{u/cj}^0|^2}\;.
\end{align}
Moreover, the quark masses are dependent on $\phi$ through the self-energy correction at loop level. The one-loop corrections to up-type (down-type) quark mass $\tilde{m}_i$ ($m_j$) are given by~\cite{Dine:2024bxv}
\begin{align}
\frac{\Delta \tilde{m}_{i}(\phi)}{\tilde{m}_{i}}&=- \frac{3}{32\pi^2}\sum_{j}|V_{ij}(\phi)|^2 \Big({m_j\over v/\sqrt{2}}\Big)^2\log\left(\frac{\Lambda_{\text{UV}}}{v/\sqrt{2}}\right)\;,
\label{mass dependence1}\\
\frac{\Delta m_{j}(\phi)}{m_{j}}&=- \frac{3}{32\pi^2}\sum_{i}|V_{ij}(\phi)|^2 \Big({\tilde{m}_i\over v/\sqrt{2}}\Big)^2 \log\left(\frac{\Lambda_{\text{UV}}}{v/\sqrt{2}}\right)\,,
\label{mass dependence2}
\end{align}
where $\Lambda_{\text{UV}}$ denotes a high-energy scale above the electroweak scale and one takes $\log(\Lambda_{\text{UV}}/(v/\sqrt{2}))\sim 1$ to get a conservative estimate~\cite{Dine:2024bxv}.

Based on the $\phi$-dependent CKM entries in Eq.~(\ref{CKM dependence}) and quark masses in Eqs.~(\ref{mass dependence1}), (\ref{mass dependence2}), we can calculate oscillating nuclear decay rate in the rest of this section.

\subsection{Mass difference between neutron and proton}

We have demonstrated that the oscillating ultralight scalar DM field $\phi$ in Nelson-Barr model induces time-dependence in both the CKM matrix elements and quark masses. Subsequent analysis will investigate how these modifications affect nuclear decay rates. Following analogous analysis in Ref.~\cite{Zhang:2023lem}, the $\phi$-induced corrections predominantly manifest through variations in the neutron-proton mass difference and nuclear binding energy. This subsection focuses specifically on the Nelson-Barr model's correction to the neutron-proton mass difference $m_n-m_p$, whose origin comprises both QED and QCD contributions. The QED contribution remains manifestly $\phi$-independent~\cite{Gasser:2020mzy}
\begin{eqnarray}
(m_n-m_p)^{\rm QED}=-(0.58\pm 0.16)~{\rm MeV}\;.
\end{eqnarray}

The QCD contribution can be derived from the next-to-leading order (NLO) pion-nucleon chiral Lagrangian and is given by~\cite{Bernard:1996gq}
\begin{align}
(m_n-m_p)^{\rm QCD}=4c_5B_0 (m_u-m_d)+\mathcal{O}(m^4_{\pi})~,
\end{align}
where $c_5=(-0.074\pm 0.006)~{\rm GeV}^{-1}$ and $B_0=|\langle 0|\bar{u}u|0\rangle|/F_\pi^2\approx 2.8~{\rm GeV}$ from Gell-Mann-Oakes-Renner relation~\cite{Bernard:1996gq} with $F_\pi$ being the pion decay constant.
Obviously, the mass difference correction arises from modifications to the $u$ and $d$ quark masses in Eqs.~(\ref{mass dependence1}) and (\ref{mass dependence2}).
After inserting the mass corrections given by $\phi$-dependent CKM matrix elements, we thus obtain
\begin{align}
m_n-m_p&=(m_n-m_p)^{\rm QED}+(m_n-m_p)^{\rm QCD}\notag\\
&\approx -0.58~{\rm MeV}+4c_5B_0{(}\tilde{m}_u+\Delta \tilde{m}_u{(}\phi {)}-m_d-\Delta m_d{(}\phi {))}\nonumber\\
&\approx (1.55-1.75\times 10^{-7} \phi/f)~{\rm MeV}\;.
\end{align}

\subsection{ Binding energy of nucleon}

The modifications to nuclear binding energy within the Nelson-Barr framework entail greater complexity, as multiple mesons contribute to binding energies. The dominant contribution arises from pion. In this subsection, we first focus on deriving the $\phi$-dependent deuteron binding energy $B_2(\phi)$. Our analysis is restricted to pion contributions and intends to obtain the $\phi$-induced modification to nuclear binding energy $\Delta B_2(\phi)$. We employ a method analogous to that in Ref.~\cite{Ubaldi:2008nf}. Starting with the pion potential $V_{\pi}(r)$, we calculate the correction to the binding energy $\Delta B_2(\phi)$ using first-order perturbation theory. Finally, based on the deuteron binding energy, we derive the $\phi$-dependent averaged binding energy of tritium $^3{\rm H}$.

According to the one-pion exchange (OPE) model, two nucleons are mediated by pion field of mass $m_{\pi}$ and the potential between two nucleons has a simple form $V_{\pi}(r)\propto-\frac{g_{\pi NN}^2}{4\pi} \frac{e^{-m_{\pi}r}}{r}$ with $g_{\pi NN}$ being pion-nucleon coupling constant. In the complex case with non-zero spin and isospin, the potential $V_{\pi}$ can be obtained from  axial dipole-dipole interaction and is given by \cite{Lee:2020tmi}
\begin{align}
\label{Vpi}
V_\pi (r)  =  (\boldsymbol{\tau_1} \cdot \boldsymbol{\tau_2}) ( \boldsymbol{\sigma_1}
\cdot\boldsymbol{\sigma_2}) \frac{g_{\pi NN}^2}{4\pi} \frac{1}{12} \left(\frac{m_\pi}{m_N} \right)^2
\frac{e^{-m_\pi r}}{r}~,
\end{align}
where $\boldsymbol{\tau_1}$ and $\boldsymbol{\tau_2}$ are the isospins of two nucleons, $\boldsymbol{\sigma_1}$ and $\boldsymbol{\sigma_2}$ are their spins, and $m_N$ denotes the mass of nucleon. We can rewrite $(\boldsymbol{\tau_1} \cdot \boldsymbol{\tau_2}) ( \boldsymbol{\sigma_1}
\cdot\boldsymbol{\sigma_2})$ by their total spin $S$ and isospin $I$ through following relations
\begin{align}
\boldsymbol{\sigma_1}\cdot\boldsymbol{\sigma_2}=2S(S+1)-3\;,~~\boldsymbol{\tau_1}\cdot\boldsymbol{\tau_2}=2I(I+1)-3\;.
\end{align}
In Nelson-Barr model, the potential $V_{\pi}(r)$ should be a function of $\phi$  because of the  $\phi$-dependence in pion  mass $m_{\pi}$, nucleon mass $m_N$ and pion-nucleon coupling constant $g_{\pi NN}$. The $\phi$-dependent $m_{\pi}$ and $m_N$ are given by~\cite{Brower:2003yx}
\begin{eqnarray}
m_\pi^2(\phi)&=&B_0(\tilde{m}_u+\Delta \tilde{m}_u(\phi)+m_d+\Delta m_d(\phi))\;,\\
m_N (\phi) &=& m_0 -4c_1 m_\pi^2(\phi) - \frac{3g_A^2m_\pi^3(\phi)}{32\pi F_\pi^2}\;,
\end{eqnarray}
where $c_1 =-1.1~{\rm GeV}^{-1}$, $g_A=1.27$, $F_\pi=92.2~\mathrm{MeV}$ and $m_0 \simeq 865~\mathrm{MeV}$~\cite{Lee:2020tmi}.
The consequent $g_{\pi NN}$ is~\cite{Fettes:1998ud}
\begin{align}
g_{\pi NN}(\phi) = \frac{g_A \, m_N(\phi)}{F_\pi} \,\left(1 - \frac{2m_\pi^2(\phi)\bar{d}_{18}}{g_A} \right)\;,
\end{align}
where $\bar{d}_{18}=-0.47~\mathrm{GeV}^{-2}$.
The above results lead to the $\phi$-dependent pion potential $V_\pi(\phi)$ and the perturbation correction $H^{\prime}=V_{\pi}\left( \phi \right) -V_{\pi}\left( \phi =0 \right)$ in Hamiltonian. We can obtain the following correction of binding energy $\Delta B_2(\phi)$ through the first-order correction of quantum perturbation theory
\begin{align}
\Delta B_2\left( \phi \right)=-\left< \psi^{(0)} |V_{\pi}\left( \phi \right) -V_{\pi}\left( \phi =0 \right) |\psi^{(0)} \right> \;,
\end{align}
where $\psi^{(0)}$ denotes the zero-order approximate wave function. To simplify the calculation,
we adopt the approach of Ref.~\cite{Ubaldi:2008nf} and approximate the ground-state wave function $\psi^{(0)}$ using a three-dimensional square well potential which is a good approximation to the potential in Eq.~(\ref{Vpi})
\begin{align}
\label{psi0}
\psi^{(0)}(r) = \left\{ \begin{array}{ll}
		A \frac{\sin kr}{r}~, & r<R \\
		B \frac{e^{-\rho r}}{r}~,& r>R
\end{array} \right.
\end{align}
where $R = 8.62 \times 10^{-3}$ MeV$^{-1}$, $k = 212$ MeV, $\rho = 46.4$ MeV, $A = 2.31$ MeV$^{1/2}$, $B = 1.44 A$.

Now, we can calculate $\left< \psi^{(0)} |V_{\pi}\left( \phi =0 \right) |\psi^{(0)} \right>$ as
\begin{align}
\left< \psi^{(0)} |V_{\pi}\left( \phi =0 \right) |\psi^{(0)} \right>&=\left( 2I\left( I+1 \right) -3 \right) \left( 2S\left( S+1 \right) -3 \right) \frac{g_{\pi NN}^2}{4\pi}\frac{1}{12}\left( \frac{m_{\pi}}{m_N} \right) ^2\times \notag\\
&\int d\Omega \Big[ \int_0^R  dr A^2 \sin^2kr \frac{e^{-m_{\pi}r}}{r}+
\int_R^{\infty}  dr B^2  \frac{e^{-2\rho r-m_{\pi}r}}{r}\Big]\nonumber\\
&=4\pi \left( 2I\left( I+1 \right) -3 \right) \left( 2S\left( S+1 \right) -3 \right) \frac{g_{\pi NN}^2}{4\pi}\frac{1}{12}\left( \frac{m_{\pi}}{m_N} \right) ^2\times\notag\\
&\Big[ A^2\Big(\frac{1}{4}\ln \left( 1+\frac{4k^2}{m_{\pi}^{2}} \right) -\frac{1}{2}E_1\left( m_{\pi}R \right) +\frac{1}{2}{\rm Re}\big( E_1( ( m_{\pi}-2ik ) R )\big ) \Big)\notag\\
&+B^2E_1 ( \left( m_{\pi}+2\rho \right)R {)} \Big ]\;,
\label{eq:Vphi=0}
\end{align}
where $S=1$, $I=0$, $m_{\pi}=139.57~{\rm MeV}$, $m_N=938.92~{\rm MeV}$, and $g^2_{\pi NN}/4\pi=13.7$~\cite{Lee:2020tmi}.
Here we use the following integrals
\begin{align}
\int^{\infty}_{0}\frac{e^{-ax}\sin^2(bx)}{x}dx&\approx\frac{1}{4}\ln (1+\frac{4b^2}{a^2})~,\\
\int^{\infty}_{R}\frac{e^{-ax}\sin^2(bx)}{x}dx&\approx\frac{1}{2}[E_1(aR)-{\rm Re}\big( E_1((a-2ib)R) \big)]\;,
\end{align}
where $E_1(kR)\equiv \int^{\infty}_{R}\frac{e^{-kx}}{x}dx$ is the first-order exponential integral function.
After replacing the masses $m_{\pi}$, $m_N$ and coupling constant $g_{\pi NN}$ in Eq.~(\ref{eq:Vphi=0}) with $\phi$-dependent quantities, we can obtain $\left< \psi^{(0)} |V_{\pi}\left( \phi  \right) |\psi^{(0)} \right>$ and finally get
\begin{align}
\Delta B_2(\phi)=-\left< \psi^{(0)} |V_{\pi}\left( \phi  \right) |\psi^{(0)} \right>+
\left< \psi^{(0)} |V_{\pi}\left( \phi =0 \right) |\psi^{(0)} \right>\simeq (0.0969 ~\phi/f) ~{\rm eV}~.
\end{align}
Then, the total binding energy of deuteron $B_2(\phi)$ is given by
\begin{align}
B_2(\phi)&=B_2+\Delta B_2(\phi)\notag\\
&=(2.22+9.69\times 10^{-8} ~\phi/f)~{\rm MeV}\;,
\end{align}
where the original binding energy $B_2\approx 2.22~{\rm MeV}$ is obtained by solving the radial Schr\"{o}dinger equation for the two-nucleon system when taking $\phi=0$ in all quantities~\cite{Lee:2020tmi}.

Next, we consider the binding energy of $^3{\rm H}$. Because SU(4) symmetry is an approximate symmetry of low-energy nuclear physics \cite{Wigner:1936dx}, the light
nuclei with up to $n=4$ nucleons follow the same universal behavior \cite
{Bedaque:1998kg,Bedaque:1998km,Bedaque:1999ve,Platter:2004he,Platter:2004zs}. Then, we can use empirical conclusion
in Ref.~\cite{Gattobigio:2012tk} that $\big(  \overline{B}_n/ {B} \big)^{1/4}-\big(  \overline{B}_2/ {B} \big)^{1/4}$ remains approximately constant, where $B$ denotes binding energy scale and is approximately equal
to the value of $\overline{B}_4$. The ``bar'' here indicates an average over states which become degenerate in the low-energy SU(4) symmetry limit. We thus have the following relation
\begin{align}
\label{Bn/B4}
\big(  \overline{B}_n(\phi)/ \overline{B}_4 \big)^{1/4}-\big(  \overline{B}_2(\phi)/ \overline{B}_4 \big)^{1/4}=
\big(  \overline{B}_n/ \overline{B}_4 \big)^{1/4}-\big(  \overline{B}_2/ \overline{B}_4 \big)^{1/4}\;,
\end{align}
where $n=3,4$. Taking $n=3$ in Eq.~(\ref{Bn/B4}), we get
\begin{equation}
\overline B_3(\phi)=\left(\overline B_2(\phi)^{1/4}+\overline B_3^{1/4}-\overline B_2^{1/4}\right)^4,
\end{equation}
where $\overline B_2\approx 1.03~{\rm MeV}$ is obtained by averaging over states including the physical deuteron and spin-singlet channel and as a result $\overline B_2(\phi)\sim (1.03+9.69\times 10^{-8} ~\phi/f)~{\rm MeV}$. $\overline B_3\approx 8.1~{\rm MeV}$ is given by averaging over states with physical $^3{\rm H}$ and $^3{\rm He}$ \cite{Zhang:2023lem}.
Finally, we obtain the $\phi$-dependent averaged  binding energy of $^3{\rm H}$
\begin{align}
\overline{B}_3(\phi)\simeq(8.1+4.55\times 10^{-7}\phi/f)~{\rm MeV}\;.
\end{align}

\subsection{Nuclear decay rate}

After establishing the modifications to the neutron-proton mass difference and nuclear binding energies in Nelson-Barr model, we now proceed to evaluate their implications for nuclear decay rates. These modifications affect decay rate $\Gamma$ primarily through the mass difference between initial and final states in $\beta$ decay process
\begin{align}
M_{i}(\phi) - M_{f}(\phi)
&= (m_n - m_p)(\phi)  + B_f(\phi)-B_i(\phi)~,
\end{align}
where $M_{i (f)}$ denotes the mass of the initial (final) nuclear state, and the individual initial and final state binding energies are given by $B(\phi)_{i/f}
\simeq B_{i/f}\frac{\overline B_3(\phi)}{\overline B_3}$ with $B_f-B_i\approx-0.76~{\rm MeV}$ from Ref.~\cite{Zhang:2023lem}. After combining previous results, we obtain
\begin{align}
M_{i}(\phi) - M_{f}(\phi)
\simeq (0.79+1.33\times 10^{-7}\phi/f)~{\rm MeV}\;.
\end{align}
The modification part induced by $\phi$ is
\begin{align}
\label{mi-mf}
\delta M_{i}(\phi) - \delta M_{f}(\phi)
\simeq (1.33\times 10^{-7}\phi/f)~{\rm MeV}\;.
\end{align}
This $\phi$-modification for mass difference between initial and final states induces the perturbation of phase space integral $I^{\beta}$ and further modifies the decay rate $\Gamma$. We start from the phase space integral
\begin{align}
\label{phase space}
I^{\beta}=\frac{1}{m_e^5}\int_{m_e}^{E_{\rm max}}F_0(Z+1,E_e)p_e E_e(E_{\rm max}-E_e)^2dE_e~,
\end{align}
where $E_e$ is electron energy, $p_e$ is electron momentum, $m_e$ is electron mass, $Z$ is atomic number, $F_0(Z+1,E_e)$ is Fermi function~\cite{Dvornicky:2011fm} and $E_{\rm max}$ is the maximum possible electron energy given by
\begin{align}
E_{\rm max}=\frac{	M_i^2+m_e^2-(M_f+m_{\nu})^2}{2M_i}
\end{align}
with neutrino mass $m_{\nu}$. After expanding $E_{\rm max}$ to the lowest order and inserting Eq.~(\ref{mi-mf}), we have
\begin{align}
\label{Emax}
E_{\rm max}(\phi)&\simeq E_{\rm max}|_{\delta M_{i/f}=0}+\delta M_i(\phi)-\delta M_f(\phi)\nonumber\\
&\simeq 0.53+1.33\times 10^{-7}\phi/f~{\rm MeV}\;.
\end{align}
Below we define the $\phi$-dependent part as $\delta E(\phi)=1.33\times 10^{-7}\phi/f~{\rm MeV}$.
Numerical integration of phase space in Eq.~(\ref{phase space}) yields
\begin{align}
\frac{\delta I^\beta(\phi)}{I^\beta}&=\frac{I^\beta(\phi)-I^\beta}{I^\beta}	 \notag\\
&=\frac{\int_{m_e}^{E_{\rm max}+\delta E(\phi)}  F_0(Z+1,E_e)p_e E_e \big(  E_{\rm max}+\delta E(\phi)-E_e \big)^2dE_e}{\int_{m_e}^{E_{\rm max}}
F_0(Z+1,E_e)p_e E_e \big(  E_{\rm max}-E_e \big)^2dE_e
} -1  \notag\\
&\simeq 0.184252 \times\left(\frac{\delta E(\phi)}{\mathrm{keV}}\right)=2.44\times 10^{-5} \phi/f~,
\end{align}
where $\delta I^\beta(\phi)$ is a small perturbation for $I^\beta$ induced by $\phi$. For the fractional change $I_0$ in the $\beta$ decay rate $\Gamma$, we have the following approximation~\cite{Zhang:2023lem}
\begin{align}
I_0(\phi)=\frac{\Gamma(\phi)-\Gamma}{\Gamma}\sim \frac{\delta I^\beta(\phi)}{I^\beta} = 2.44\times 10^{-5} \phi/f\;.
\end{align}
In experiments, the residual $I(\phi)$ of decay rate $\Gamma$ is defined as
\begin{align}
\label{Iphi}
I(\phi)&=\frac{\Gamma(\phi)-\langle \Gamma(\phi)\rangle}{\langle \Gamma(\phi)\rangle}=\frac{I_0(\phi)+1}{\langle I_0(\phi)\rangle+1}-1\notag\\
&=2.44\times 10^{-5} \frac{\sqrt{2\rho _{\text{DM}}}}{fm_{\phi}}\cos \left( m_{\phi}t +\delta \right)\;,
\end{align}
where $\langle \Gamma(\phi)\rangle =\big(\langle I_0(\phi)\rangle+1\big)\times \Gamma$ is the time average of $\Gamma(\phi)$ with $\langle I_0(\phi)\rangle \propto \langle \cos(m_{\phi}t +\delta)\rangle=0$, and $\delta$ is a random phase.

\section{Constraint on Nelson-Barr dark matter}
\label{sec:Cons}

The results from the previous section demonstrate that the ultralight scalar DM field $\phi$ in Nelson-Barr model induces oscillating residual $I(\phi)$ in tritium decay rate.
This enables experimental detection of DM signatures through time-dependent spectral residuals. Ref.~\cite{Zhang:2023lem} analyzed
the 12 years of data for the decay of
$^3{\rm H}$ provided by
European Commission’s Joint Research Centre
(JRC) at the Directorate for Nuclear Safety and Security in Belgium~\cite{Pomme:2016ckl}.
However, the existing measurements exhibit no statistically significant periodic oscillation in $^3{\rm H}$ decay rate upon hypothesis testing.
This null observation can be translated into the constraint on Nelson-Barr model parameters by using a standard frequentist approach illustrated in Ref.~\cite{Centers:2019dyn}.

In this section, we quantify the Type-I error $\alpha$ and Type-II error $\beta$ within a hypothesis testing framework by taking the phase-space modulation of decay rates derived in Eq.~(\ref{Iphi}). For a well-designed detector with high sensitivity, the condition $\beta\leq \alpha$ naturally applies. Consequently, this requirement places systematic exclusion bounds on the DM mass $m_{\phi}$ and its decay constant $f$ in Nelson-Barr model.

\subsection{Frequentist approach}

The hypothesis testing is usually used to verify theoretical models and eliminate false positive results caused by noise. In hypothesis testing, the p-value is computed to assess whether the experimental data are consistent with the null hypothesis. A common null hypothesis is that any observed positive signal arises purely from statistical fluctuations or noise. For instance, the analysis in Ref.~\cite{Zhang:2023lem} reports a p-value greater than 0.05 for the experimental data on the
$^3{\rm H}$ decay rate. Thus, at the 95\% confidence level (CL), no statistically significant periodicity is observed in the decay rate.

After confirming that there is no significant periodicity in $^3{\rm H}$ decay rate, we can follow Ref.~\cite{Centers:2019dyn} to adopt frequentist approach to place constraints on the Nelson-Barr model parameters.
The frequentist approach requires us to first obtain Type-I error $\alpha$ and Type-II error $\beta$~\cite{2010All}.
The Type-I error $\alpha$ represents the probability of falsely detecting a signal (e.g., DM signal) when in fact the positive result of an experiment is caused by noise.
Type-II error $\beta$ is the possibility of failing to detect a true signal, but incorrectly attributing it to noise. To calculate $\alpha$, we need to integrate the likelihood function $\mathcal{L}(A|A_s=0)$ of the null hypothesis (where $A_s=0$ indicates no true signal exists) over the region exceeding the detection threshold $A^{\rm dt}$~\cite{Centers:2019dyn}
\begin{align}
\label{alpha}
\alpha=\int_{A^{\rm dt}}^{\infty} \mathcal{L}(A|A_s=0)dA\;,
\end{align}
where $A=|\tilde{d}_p|/\sqrt{N\sigma^2}$ and $A_s=|\tilde{s}_p|/\sqrt{N\sigma^2}$, $\tilde{d}_p$ ($\tilde{s}_p$) is the discrete Fourier transformation (DFT) component of experimental data $d(t)$ (theoretically assumed signal $s(t)$)~\cite{Centers:2019dyn}, $N$ is the number of data points, $\sigma$ is standard deviation of data.
The detection threshold $A^{\rm dt}$ is determined by the confidence level ${\rm CL}=1-\alpha$. For the standard 95\% CL ($\alpha=0.05$), we obtain $A^{\rm dt}\approx 1.73$ by solving the integral equation Eq.~(\ref{alpha}).
The Type-II error rate $\beta$ is calculated by integrating the likelihood function $\mathcal{L}(A|A_s\neq 0)$ of the alternative hypothesis over the region below the detection threshold $A^{\rm dt}$~\cite{Centers:2019dyn}~\footnote{Note that in Ref.~\cite{Centers:2019dyn}, the Type-II error is defined as $1-\beta$. Here we follow the definition in textbook~\cite{1998Statistical} and define it as $\beta$.}
\begin{align}
\label{beta}
\beta=\int^{A^{\rm dt}}_0 \mathcal{L}(A|A_s\neq 0)dA\;,
\end{align}
where $A_s\neq 0$ means that  the observed signal consists of both DM signal $s(t)$ and background noise.
For a well-designed experiment, the Type-II error rate $\beta$ (probability of missing a true signal) should typically be smaller than the Type-I error rate $\alpha$ (probability of false detection).

The theoretically assumed signal $s(t)$ in this  experiment is residual $I(\phi)$ and the noise is Gaussian. According to the central limit theorem~\cite{2004Introduction}, the distribution of DM signal is of the following form~\cite{Centers:2019dyn}
\begin{align}
\label{L signal}
\mathcal{L}(\tilde{d}_p|\tilde{\Phi},f,\delta)=\frac{1}{\pi N \sigma^2} {\exp}\Big( -\frac{|\tilde{d}_p-\tilde{I}_p|^2}{N\sigma^2} \Big)~,
\end{align}
where $\tilde{\Phi}$ denotes the amplitude of  field $\phi$, $\tilde{I}_p$ is the DFT component of  $I(\phi)$ and $\delta$ is the random phase of $I(\phi)$. Eq.~(\ref{L signal}) represents the possibility of obtaining the experimental data $\tilde{d}_p$ when the theoretical parameters $f$, $\tilde{\Phi}$ and random phase $\delta$ are fixed.
One can rewrite
Eq.~(\ref{L signal}) in polar coordinates \cite{Centers:2019dyn}
\begin{align}
\mathcal{L}(\tilde{d}_p|\tilde{\Phi},f,\delta)=\frac{1}{\pi N \sigma^2} {\exp}\Big( -\frac{|\tilde{d}_p|^2+|\tilde{I}_p|^2}{N\sigma^2} \Big)
{\exp}\Big( \frac{2|\tilde{d}_p||\tilde{I}_p|\cos(\phi-\delta)}{N\sigma^2} \Big)\;,
\end{align}
where $\tilde{d}_p=|\tilde{d}_p|e^{i\phi}$, $\tilde{I}_p=|\tilde{I}_p|e^{i\delta}$ and $\phi$ denotes the phase of $\tilde{d}_p$. $\mathcal{L}(|\tilde{d}_p|~|\tilde{\Phi},f,\delta)$ is then given by integrating out phase $\phi$
\begin{align}
\mathcal{L}(|\tilde{d}_p|~|\tilde{\Phi},f,\delta)=\int_{0}^{2\pi}d\phi \frac{1}{\pi N \sigma^2} {\exp}\Big( -\frac{|\tilde{d}_p|^2+|\tilde{I}_p|^2}{N\sigma^2} \Big)
{\exp}\Big( \frac{2|\tilde{d}_p||\tilde{I}_p|\cos(\phi-\delta)}{N\sigma^2} \Big)\;.
\end{align}
To make a variable replacement $|\tilde{d}_p|\rightarrow A$, one can multiply the above $\mathcal{L}(|\tilde{d}_p|~|\tilde{\Phi},f,\delta)$ by a delta function  $\delta(A-\frac{|\tilde{d}_p|}{\sqrt{N\sigma^2}})$ and integrate over $|\tilde{d}_p|$
\begin{align}
\mathcal{L}(A|A_s)&=\int d|\tilde{d}_p|\cdot|\tilde{d}_p|\delta(A-\frac{|\tilde{d}_p|}{\sqrt{N\sigma^2}}) \mathcal{L}(|\tilde{d}_p|~|\tilde{\Phi},f,\delta)\notag\\
&=\int d|\tilde{d}_p| \delta(A-\frac{|\tilde{d}_p|}{\sqrt{N\sigma^2}}) \int_{0}^{2\pi}d\phi \frac{|\tilde{d}_p|}{\pi N \sigma^2} {\exp}\Big( -\frac{|\tilde{d}_p|^2+|\tilde{I}_p|^2}{N\sigma^2} \Big)
{\exp}\Big( \frac{2|\tilde{d}_p||\tilde{I}_p|\cos(\phi-\delta)}{N\sigma^2} \Big)\notag\\
&=\int_0^{2\pi}d\phi \frac{A}{\pi} e^{-A^2-\frac{|\tilde{I}_p|^2}{N\sigma^2}} e^{2A
 \frac{|\tilde{I}_p|}{\sqrt{N\sigma^2}} \cos(\phi-\delta)}  \notag
\\
&=2{A} e^{-A^2-A_s^2}\mathcal{I}_0(2A A_s)\;,
\label{eq:Las}
\end{align}
where $\mathcal{I}_0(x)=\frac{1}{\pi}\int_0^{\pi} e^{x\cos \phi} d\phi$ is the modified Bessel function of the first kind.
After taking $A_s=0$ in Eq.~(\ref{eq:Las}), one can obtain the likelihood in Eq.~(\ref{alpha}).

\subsection{Stochastic amplitude correction}

Before calculating the error rates $\alpha$ and $\beta$, we need to consider the amplitude $\tilde{\Phi}$ more carefully.
The amplitude $\tilde{\Phi}$ is usually assumed to be a constant $\Phi_{\rm DM}=\sqrt{2\rho_{\rm DM}}/m_{\phi}$. This is a good approximation when the observation time of the experiment $T_e$ is much longer than the coherence time $\tau$ of the $\phi$ field. However, when the mass $m_{\phi}$ is less than $10^{-16}~{\rm eV}$, the coherence time $\tau$ is greater than 1 year~\cite{Centers:2019dyn}. As a result, the averaged observation time of $^3{\rm H}$ experiment $T_e=1~{\rm second}$~\cite{Zhang:2023lem} or $1~{\rm hour}$~\cite{Zhang:2023lem,Bikit:2013dna} is less than $\tau$. Thus, the fixed field amplitude $\tilde{\Phi}=\Phi_{\rm DM}$ is not a good approximation anymore.

In this low-mass regime, DM is regarded as virialized ultralight fields (VULFs) which are composed of a large number of individual modes with randomized phases~\cite{Centers:2019dyn}. The amplitude $\tilde{\Phi}$ which depends on the experimental observation is not a fixed average value, but a series of random values $\Phi_0$. Each measurement gives a random value of $\Phi_0$, and they satisfy the Rayleigh distribution~\cite{Foster:2017hbq,Centers:2019dyn}
\begin{align}
p({\Phi_0}) =
\frac{{2{\Phi _0}}}{{\Phi _{{\rm{DM}}}^2}}\exp \left( { - \frac{{\Phi _0^2}}{{\Phi _{{\rm{DM}}}^2}}} \right)\,.
\end{align}
This stochastic nature of amplitude leads to the correction to likelihood function Eq.~(\ref{L signal})
\begin{align}
\mathcal{L}(\tilde{d}_p|{\Phi_0},f,\delta)\to \mathcal{L}(\tilde{d}_p|{\Phi_0},f,\delta)p(\Phi_0)\;.
\end{align}
In order to consider the contributions to likelihood function from all the random amplitudes $\Phi_0$, it is necessary to marginalize $\Phi_0$ by performimg the following integration
\begin{align}
\label{L stochastic}
\mathcal{L}(\tilde{d}_p|f,\delta)=
\int_0^{\infty}d\Phi_0 \mathcal{L}(\tilde{d}_p|{\Phi_0},f,\delta) p(\Phi_0)\;.
\end{align}
Similarly, after rewriting it by $A$ and $A_s$, we obtain the new likelihood function
\begin{align}
\label{L as 2}
\mathcal{L}(A|A_s)=\int_0^{\infty}   d\Phi_0  {2}Ae^{-A^2-A_{s}^{2}}\mathcal{I}_0\left( 2AA_s \right)p\left(\Phi_0\right) \;.
\end{align}

\subsection{Constraint on Nelson-Barr DM parameters}

Now, we are ready to obtain the constraint by applying the condition $\alpha\geq\beta$. First, we calculate the DFT component $\tilde{I}_p$ of the signal $I(\phi)$. After discretizing time $t$ into $t=\frac{n}{N}T_{\rm total},~n=0,1,\cdots,N-1$ with $T_{\rm total}$ being the total running time of experiment and $N$ the total number of data, we can get the DFT component of $I(t)$ as
\begin{align}
\tilde{I}_p&=\sum_{n=0}^{N-1} I(t)\cdot e^{-i\frac{2\pi n}{N}p}\notag\\
&=  2.44\times 10^{-5}\frac{\Phi_0}{f} \sum_{n=0}^{N-1} \Big(\frac{e^{i\delta}}{2}e^{-i\frac{2\pi n}{N}(p-\frac{m_{\phi}}{2\pi}T_{\rm total})}+
\frac{e^{-i\delta}}{2}e^{-i\frac{2\pi n}{N}(p+\frac{m_{\phi}}{2\pi}T_{\rm total})}
\Big)\notag\\
&=2.44\times 10^{-5}\frac{N}{2}\frac{\Phi_0}{f}e^{i\delta}\;,
\end{align}
where the index $p$ denotes the DFT frequencies $f_p=p/T_{\rm total}=m_\phi/(2\pi)$.

\begin{figure}
\centering
\includegraphics[width=0.9\linewidth]{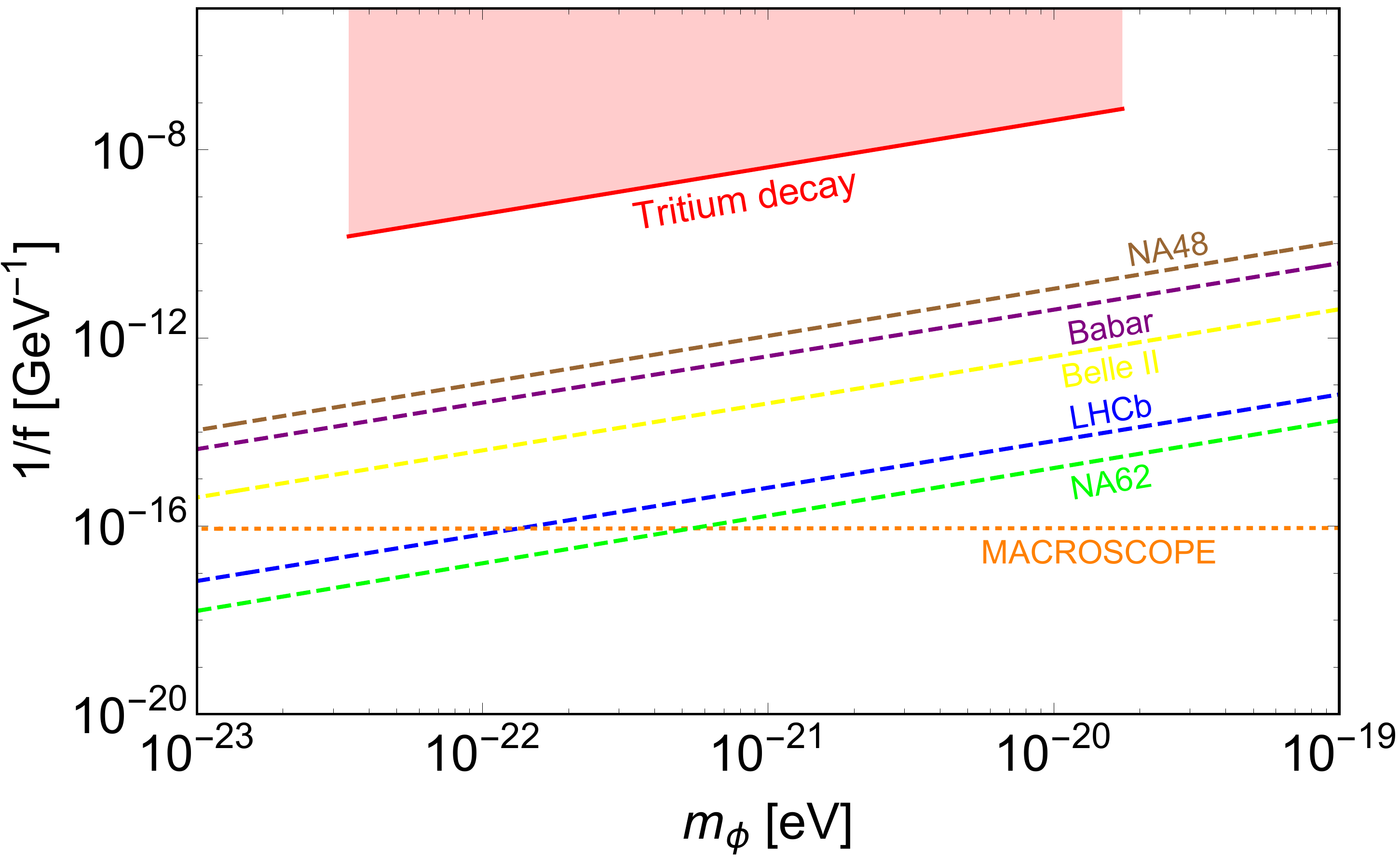}
\caption{The 95\% CL exclusion limit on Nelson-Barr DM decay constant $1/f$ from the non-observation of periodic variations in $^3{\rm H}$ decay data (red region). The constraints from other experiments are also shown for comparison, including the current MICROSCOPE~\cite{Touboul:2017grn,MICROSCOPE:2022doy} (orange dotted line), and the potential reaches from the measurement of CKM matrix element variation in experiments such as NA62~\cite{NA62:2017rwk} (green dashed line), NA48~\cite{NA48:2006jeq} (brown dashed line), Babar~\cite{delRe:2003if} (purple dashed line), Belle II~\cite{Belle-II:2024xwh} (yellow dashed line) and LHCb~\cite{LHCb:2024jpt} (blue dashed line).
}
\label{Constraint}
\end{figure}

Then, $A_s$ in likelihood Eq.~(\ref{L as 2}) becomes
\begin{align}
\label{Asth}
A_s=\frac{|\tilde{I}_p |}{\sqrt{N\sigma ^2}}=2.44\times 10^{-5}\frac{\sqrt{N}}{2\sigma}\frac{\Phi _0}{f}\;.
\end{align}
After inserting Eq.~(\ref{L as 2}) into Eqs.~(\ref{alpha}) and (\ref{beta}), we finally obtain
\begin{align}
\label{a>b2}
\alpha=\int_{1.73}^{\infty}dA~ 2A e^{-A^2}\geq\beta=\int_0^{1.73}dA\int_0^{\infty}dA_s2Ae^{-A^2-(A_{s})^2}\mathcal{I}_0\left( 2AA_s \right) bA_se^{-b(A_{s})^2/2}~,
\end{align}
where the coefficient $b$ is defined as
\begin{align}
\label{a}
b=\frac{4\left( m_{\phi}f \right) ^2\sigma ^2}{\left( 2.44\times 10^{-5} \right) ^2\rho _{\text{DM}}N}
\end{align}
with $N=50000$~\cite{Zhang:2023lem} and $\sigma=0.4\%$~\cite{Pomme:2016ckl}. The numerical integration of Eq.~(\ref{a>b2}) results in $b\geq 0.0349$. We then obtain the constraint on Nelson-Barr DM parameters according to Eq.~(\ref{a}).
The experimental data of $^3{\rm H}$~\cite{Zhang:2023lem} show the frequency range is $8.0 \times 10^{-9}-4.6\times 10^{-6} ~{\rm Hz}$. The corresponding range of DM mass is $3.4\times 10^{-23}~{\rm eV}<m_\phi<1.7\times 10^{-20}~{\rm eV}$.
Finally, we derive the 95\% CL exclusion limit on Nelson-Barr model parameters shown by the red line in Fig.~\ref{Constraint}. Decay constant $f$ below $7.0\times 10^{9}-1.4\times 10^7$ GeV is excluded for mass $m_\phi$ in the range of $3.4\times 10^{-23}-1.7\times 10^{-20}~{\rm eV}$.
The constraints from other experiments are also shown for comparison, including the current MICROSCOPE~\cite{Touboul:2017grn,MICROSCOPE:2022doy} and the potential reaches from the measurements of CKM matrix element variations in other experiments~\cite{Dine:2024bxv}.

Finally, we compare our results in Nelson-Barr model with that of axion model in Ref.~\cite{Zhang:2023lem}. Note that the Taylor series expansion of Eq.~(\ref{eq:V2}) results in the linear dependence of $\phi/f$ in the Eq.~(\ref{CKM phi}). Thus, the decay rate residual in Eq.~(\ref{Iphi}) linearly depends on $1/f$, while the residual in axion model is proportional to $1/f_a^2$~\cite{Zhang:2023lem} with $f_a$ being the decay constant of axion. On the other hand, as shown in the one-loop correction of quark masses in Eqs.~(\ref{mass dependence1}) and (\ref{mass dependence2}), the factor of $(m_q/v)^2$ leads to significant suppression of the mass corrections, as well as the correction of neutron-proton mass difference, the correction of pion mass, nucleon mass and pion-nucleon coupling coefficient. The correction of nuclear binding energy also becomes small. For the axion model, there is no such suppression. As a result, if one takes $m_\phi=m_a=10^{-21}$ eV and $f=f_a=10^{13}$ GeV for illustration, the residual $I(\theta)$ in axion model given by Eq.~(12) in Ref.~\cite{Zhang:2023lem} is four orders of magnitude greater than that in the Nelson-Barr model. These distinctions induces different likelihood functions. We showed the error rates in Eq.~(\ref{a>b2}) with the coefficient $b$ for Nelson-Barr model. For axion, however, we find
\begin{eqnarray}
\beta=\int_0^{1.73}dA\int_0^{\infty}dA_s2Ae^{-A^2-(A_{s})^2}\mathcal{I}_0\left( 2AA_s \right) b' e^{-b' A_{s}}~,
\end{eqnarray}
where the coefficient $b'$ is obtained by inserting the axion residual $I(\theta)$ in the Eq.~(12) of Ref.~\cite{Zhang:2023lem}
\begin{eqnarray}
b'=\frac{2\left( m_a f_a \right) ^2\sigma}{375.2 \rho_{\text{DM}} \sqrt{N}}\geq 0.0331\;.
\end{eqnarray}
One can see that both of these two constraining results quadratically depend on $f$ or $f_a$. As a result, we can quantitatively reproduce the limit on $f_a$ ($9\times 10^{12}-1\times 10^{10}$ GeV) in Ref.~\cite{Zhang:2023lem}. We find the limit on $f$ in Nelson-Barr model is three orders of magnitude less stringent than that on $f_a$ in axion model.

\section{Conclusion}
\label{sec:Con}

Many experiments have reported periodic variations in radioactive decay rates. This periodicity may attribute to the presence of a periodic DM candidate. Nevertheless, the evidence of this annual periodicity was challenged by some more recent measurements and statistical analysis which excluded modulations of the decay rates in tritium and several heavier radioisotopes. The null significant observation of periodic variations in nuclear decay rates allows us to place constraints on potential time-varying DM.

In this work, we investigate the impact of time-dependent tritium $\beta$-decay rate on the ultralight scalar DM in Nelson-Barr model. The light scalar DM field in this framework induces periodic modulations of both CKM matrix elements and quark mass parameters. These modulations generate corresponding periodic perturbations in the neutron-proton mass difference and nuclear binding energies. Consequently, the ultralight scalar DM field results in oscillating residuals in nuclear decay rates. We implement a frequentist hypothesis-testing protocol on precision measurements of tritium $\beta$-decay rates and derive novel constraints on the DM mass $m_{\phi}$
and decay constant $f$. We find that decay constant $f$ below $7.0\times 10^{9}-1.4\times 10^7$ GeV is excluded at 95\% CL for mass $m_\phi$ in the range of $3.4\times 10^{-23}-1.7\times 10^{-20}~{\rm eV}$.

\acknowledgments

T.~L. is supported by the National Natural Science Foundation of China (Grant No. 12375096, 12035008, 11975129).

\bibliography{refs}

\end{document}